# Niels Bohr as a Physicist of Principle

Mauro Dorato (corresponding author)

Department of Philosophy, Communication and Media Studies
Roma Tre University
Via Ostiense 234
00146 Rome, Italy
mauro.dorato@gmail.com
ORCID: 0000-0002-8313-6362

&

Jan Faye
Professor Emeritus, Department of Communication
University of Copenhagen
Karen Blixens Plads 8
2300 København S
Denmark
faye@hum.ku.dk
ORCID: 0000-0002-4438-7564

## Abstract

Our main claim is that Bohr adopted a principle-theoretic approach to quantum mechanics in order to reconcile two principles: the distinction principle, which asserts the necessity of a clear epistemic distinction between a classical apparatus and a quantum system, and the non-separability principle, which asserts that the two systems are ontologically non-separable in any measurement process. The latter principle is a consequence of Bohr's argument that, during a measurement interaction, the quantum system and the measurement apparatus form a 'new kind of individuality'. To support our argument, we first demonstrate the overlooked fact that, according to Bohr, the measurement interaction is a physical, irreversible process and not merely an epistemic acquisition of new information. In this respect, quantum mechanics must be regarded not only as a universal theory, but also as revealing no non-arbitrary difference between a quantum realm and a classical realm; the distinction is purely pragmatic. The tension between the epistemic necessity to rely on a neatly distinguishable classical realm and the ontic non-separability of the quantum system and the classical apparatus inclined Bohr to refuse to provide a constructive account of the measurement process, such as that later proposed, for instance, by spontaneous dynamical reduction models.

In his valuable book *Einstein*, Thomas Ryckman (2017) has claimed with highly plausible arguments that Einstein's philosophy of physics consisted "in the advance of, or critique of theory through the use, and presumed general validity, of certain principles; physical, formal, methodological, and metaphysical" (Ryckman 2017, p. 4). The principles to which Ryckman is referring are the light principle and the relativity principle for the special theory of relativity, the equivalence principle for the general theory of relativity, and the locality principle and the separability principle for quantum mechanics.

In this paper, we argue that Bohr's approach to the measurement process was, to a significant extent, inspired by Einstein's conviction that, at least in crucial cases of the history of science, there cannot be any *constructive* account of physical processes, and *principle-theory* approaches are – at least temporarily – not only more viable but also the only way to go (Einstein 2002).[1] Our main claim is that it is very plausible to attribute to Bohr a principle-theory approach to resolve a *conflict* between, on the one hand, Bohr's need for a clear *epistemic* distinction between a classical apparatus and a quantum system (what we will refer to as the *distinction principle*) and, *on the other*, his explicit recognition that in any measurement process the apparatus and the system are *ontologically* non-separable (the non-separability principle). The latter principle is a consequence of the fact that, according to Bohr, during a measurement interaction, the quantum system and the measurement apparatus form a "new kind of individuality" by becoming non-separable. Although Bohr never used the term 'entanglement', he had, as we will see, already clearly understood its novelty in 1927 (see Howard 1994 and Dieks 2017).[2]

The paper is divided into four logically related sections. To defend the *ontological* character of the non-separability principle, in the first section we argue in favor of the somewhat neglected point that, according to Bohr, the measurement interaction is a *physical*, irreversible process and therefore *not merely an epistemic acquisition* of new information. As is well-known, this epistemic view is typically defended by the QBists, insofar as their approach to the quantum measurement involves simply updating our subjective beliefs on the basis of the acquisition of new information (see, among many others, Brukner 2018; Brukner and Zeilinger 2003; Fuchs, Mermin, and Schack 2014). In our view, according to Bohr, the acquisition of new information is a *consequence* of a physical process.

In the *second* section, by carefully commenting on his Como paper (Bohr 1927) and by referring to other relevant papers, we shall present and discuss in detail the conflict between the two principles mentioned above. Here we discuss the sense in which Bohr's contextual theory of measurement and relational view of quantum entities enabled him to propose a solution to the conflict between the distinction principle and the non-separability principle. Bohr's opposition to a subjective intervention of the human consciousness in the measurement problem clearly shows that, against an interpretation that is now beginning to dwindle, Bohr was ready to accept the idea that – even if relative to given, well-specified experimental contexts – any *classical, macroscopic apparatus could be treated as a*

---

[1] As argued by Brown (2005), Einstein considered the principle-theory approach to the special theory of relativity only as temporary.
[2] As is well-known, the term entanglement was explicitly introduced in print by Schrödinger (1935).

*quantum object* (Howard 1994, Dieks 2017).[3] According to Bohr, not only must quantum mechanics be regarded as a universal, fundamental theory, but also no non-arbitrary difference can be said to exist between a quantum realm and a classical realm.

In light of the previous two sections, in the third we claim that Bohr's approach to the measurement problem can best be described by attributing him the claim that, due to a distinction between a *material* and a *functional* type of non-separability, quantum mechanics should be considered a principle-theory in the sense of Einstein (see also Dorato 2017). As alternative proposals to solve the conflict, in the fourth section we will critically evaluate and compare Howard's (2021), Zinkernagel's (2015), and Dieks's (2017) interpretations of Bohr's view of the classical/quantum distinction by arguing that their approaches are less convincing than ours. We conclude by showing why the tension between, on the one hand, the *epistemic* necessity to rely on a neatly distinguishable classical realm, and, on the other, the *ontic* non-separability of the quantum system and the classical apparatus inclined Bohr to refuse to give a constructive account of the measurement process of the kind later proposed by spontaneous dynamical reduction models (for instance, see Ghirardi, Rimini and Weber 1986).

## 1. Bohr's *physical* account of measurement

In a famous passage, Bohr claims that measurements are possible due to "suitable *amplification* devices with irreversible functioning such as, for example, permanent marks on a photographic plate, or to similar *practically irreversible* amplification effects like the building of a water drop around an ion in a cloud-chamber." (Bohr 1954, p. 73, emphasis added). The amplification devices are classically describable physical instruments that record observable, irreversible traces left on the classical apparatuses by quantum systems. This important point deserves two remarks.

The *first* is that Bohr's stress on the necessary presence of *irreversible* physical processes for a measurement to be possible is a response that he could have given to John Bell's later complaint that – apart from the *application* of the theory – words like 'reversible and irreversible' "should have no place in a *formulation* with any pretension to physical precision" (Bell 1990, p. 19, italics in the original). To be charitable toward Bell, however, his point should be correctly interpreted since, of course, he would not have denied that a more precise or detailed, constructive description of the nature of the measurement interaction might even require a *nomological* form of irreversibility of the kind that, according to Albert (2003), is introduced by the GRW (1986) modification of the linearity of Schrödinger's equation. In other words, John Bell probably would not have opposed the remark that – given that von

[3] This thesis has been very recently reformulated by attributing Bohr a belief in a quantum universalism not unlike that of von Neumann's (Laudisa 2024). A discussion of the conflict between Laudisa's thesis and Zinkernagel's (2015) claim that Bohr was *not* a quantum fundamentalist in Laudisa's sense must be left to another paper.

Neumann's formulation of quantum mechanics contrasts *two* distinct temporal evolutions for quantum systems, namely an irreversible and stochastic process 1 (of collapse) and a time-symmetric deterministic process 2 governed by Schrödinger's equation (von Neumann 1932, p. 186) – the distinction between reversible and irreversible physical processes has a fundamental role, since it points to a possible internal conflict of the theory. With respect to this point, Bohr and Bell would have agreed: the difference in their attitude amounts to the question of whether a constructive, more detailed or precise account of the irreversible process involved in a measurement is needed or not.

The *second* remark is that, unlike the main philosophical stance adopted by QBists, the concept 'information' for Bohr did not play any role in his description of the measurement processes: the increase in our information due to measurements is an *effect* of such physical processes. To the extent that 'information' essentially involves a *semantic* aspect – 'to be informed about *p*' entails the existence of someone's belief whose content is the true proposition *p* – it also presupposes a conscious possessor of information about the corresponding given state of affairs.[4] From 1927 onward, this dismissive attitude toward the concept of information was a major difference between Bohr's and Heisenberg's interpretations of quantum mechanics (Howard 2021, p. 522).[5] Additional evidence in favor of our attributing to Bohr a physical, non-purely informational account of the measurement process is provided by the following four points.

1 Unlike the QBists' approaches to measurements and quantum theory in general, Bohr always defended an *objectivist* account of quantum probability, where Born's rules are the result of a long series of experiments. This implies that he would have been against subjectivist/Bayesian accounts of the probability of measurement outcomes in terms of *credences* (Fuchs, Mermin, and Schack 2014; Mermin 2017, to name two important papers). By connecting objective chance (physical probability) with rational credences (degrees of belief), it would be possible to invoke Lewis' principal principle (Lewis 1980) and therefore align a subjectivist reading of quantum probabilities with Bohr's position. However, if we accept the reasonable claim that statistical laws are derived from objective frequencies, the following quotation provides strong evidence for the fact that Bohr defended an *objectivist* view of probability: "In the treatment of atomic problems, actual calculations are most conveniently carried out with the help of a Schrödinger state function, from which the *statistical laws* governing observations obtainable under specified conditions can be deduced by definite mathematical operations" (Bohr 1963, p. 5, our emphasis). Given his attribution of a purely symbolic, non-representational role to the wave function, the following quotation provides additional evidence for our claim: "The entire [quantum] formalism is to be considered as a *tool* for deriving

[4] Here we avoid discussing the controversial issue of whether information can be regarded as a *physical* notion (Adlam and Rovelli 2023), given that in most QBists' approaches, information is regarded, correctly in our opinion, as an *epistemic* notion. See Timpson (2013).
[5] Landau and Lifshitz, two authoritative physicists who wrote an important, widely read textbook on quantum mechanics, expressed the same Bohr-derived concept in a clear way: "By measurement, in quantum mechanics, we understand any process of interaction between classical and quantum objects, *occurring apart from and independently of any observer*." (quoted in Bell 1990, p. 21, our emphasis)

predictions of definite or *statistical character*, as regards information obtainable under experimental conditions described in classical terms" (Bohr 1998, pp. 143–144, our emphasis).[6] A few remarks are appropriate here. First, it is clear from these passages that for Bohr the wave function is *not* something that encodes only subjectivist, updatable information possessed by observers and used by them to make predictions, but it also enables us to make objective probabilistic predictions because it refers to objective, physical measurement processes.

2

Second, and similarly to pragmatist accounts of quantum probabilities (Healey 2012, Healey 2017), Bohr's probabilities *do represent* real properties of the world but, unlike such accounts, they do *not* refer directly to what *agents* can do with them: it is thanks to the descriptive force of such probabilities that agents can successfully use them. Considering Bohr's approval of pragmatism (Folse 2017, Murdoch 1987), however, on this point he and Healey could agree. The expression 'tool for predicting' must be explained carefully, given that Bohr's attitude toward the wave function is quite controversial (see note 9). By using "tool for predicting", he refers to his *symbolic* understanding of the quantum formalism, where 'symbolic', as in Heisenberg (1927), Dirac (1930), and Cassirer (1936),[7] is meant to oppose the view that the wave function can be regarded as a sort of *visualizable representation of the quantum state* (Chevalley 1994, and Beller 1999, p. 180). The problem of the visualizability (*Anschauunglichkeit*) of the quantum processes (Miller 1984) was of central concern for the creators of quantum mechanics, and became important in particular after Bohr's 1913 model of the atom and Heisenberg's formulation of quantum mechanics in terms of matrices, a formalism which, more than Schrödinger's, is not amenable to visualization (Beller 1999, Camilleri 2009, Cassirer 1936).[8] The problem was in part an inheritance of Kant's philosophy, according to which, roughly speaking, every physical process had to be *visualizable* as a consequence of the possibility of giving a spatiotemporal description of physical reality in terms of the two a priori forms of our intuitions, space and time.[9]

In addition to its intrinsic interest, Bohr's claim that in well-defined experimental contexts, measurements of quantum systems do not reveal preexisting magnitudes shows that he did not adhere to a strict neopositivist verification theory of meaning: reference to entities that are not directly observable for him does *not* require a reductionist account of theoretical entities in terms of sense data or observations. In this sense, following Psillos's (1999) semantic characterization of scientific realism, Bohr explicitly endorses entity realism about quantum systems. According to a strict verificationist theory of meaning, instead, there would be *no difference* between the claim that (i) there are definite magnitudes before measurement that cannot be measured due to a "disturbance" in Heisenberg's sense, and the claim that (ii) it is the interaction with the quantum system that determines the acquisition of a definite value. For a strict neopositivist, in *both cases* it would be *meaningless* to talk about any sort of magnitude without a measurement.

---

[6] In this quotation, note that 'information' is obtainable from a physical process.
[7] For Dirac's symbolic understanding of the quantum formalism in relation to Cassirer, see Ryckman (2018).
[8] This question is central to understanding Schrödinger's polemical attitude toward matrix mechanics.
[9] It also helps to explain Schrödinger's polemical attitude toward 'quantum jumps' (Schrödinger 1952).

On the contrary, if one distinguishes between them as Bohr did by embracing only (ii), one is not a neopositivist as defined above. Moreover, Bohr's opposition to the view that we measure a pre-existing value is obvious from his well-known view that the experimental set-up *defines* the conditions under which a value can be unambiguously ascribed to a quantum object: "the distinction between the objects under investigation and the measuring instrument*s* which serve *to define*, in classical terms, the conditions under which the phenomena appear." (Bohr 1958, p. 50, our emphasis). First, Bohr explicitly stated in his outline of his discussion with Einstein that "I advocated the application of the word *phenomenon* exclusively to refer to the observations obtained under specified circumstances, including an account of the *whole experimental arrangement*" (Bohr 1949, p. 238, our emphasis).[10] The meaning is, indeed, that the specified circumstances also contain a specification of the entire experimental set-up that allows us to use one or the other of the classical concepts. Second, and relatedly, 'define' here must be understood in the right way: the word refers to the fact that, for Bohr, a definite value can be assigned to a quantum system S only relative to a given, non-separable measurement apparatus and does not refer to any nominal or purely semantic definition. The selection of an apparatus determines the kind of concept it is meaningful to apply to the outcome, and the actual measurement determines the value corresponding to this concept. For Bohr, a phenomenon is the *amplified effect of the non-controllable interaction* between a quantum system and an instrument, and therefore it does not inform us about any pre-existing value. So the phenomenon depends both ontologically and semantically on the experiment that determines the conditions under which it make sense to ascribe a property to a quantum system in term of a specific classical concept. Bohr's opposition to Heisenberg's theory of measurement as 'disturbance' yields another argument in favor of the idea that measurements for him are real physical processes. Bohr's archenemy, perhaps not surprisingly, endorsed the importance of such an opposition: "When it is said that something is "measured" it is difficult not to think of the result as referring to some preexisting property of the object in question. *This is to disregard Bohr's insistence that in quantum phenomena the apparatus as well as the system is essentially involved*" (Bell 1990, ibid., our emphasis). The correct claim that, prior to measurements, in certain circumstances, quantum systems lack definite properties according to Bohr entailed that the acquisition of definite ones is due to some *physical interaction* between two real systems that, as we will see, does *not* cry out for a constructive explanation. This claim is consistent with the well-known fact that, according to Bohr, there is no unambiguous way of referring to properties of quantum systems independently of a measurement context.

[10] We thank the referee for suggesting that we place greater emphasis on Bohr's technical term 'phenomenon'.

3 Bell, however, moved a related charge against Bohr that enables us to defend our claim from a different angle. Given that Bohr often insisted on the fact that any experimental report had to be expressed in the language of classical physics, one may ask whether Bell's allegation against Bohr that, "the apparatus should not be separated off from the rest of the world into black boxes, as if it were not made of atoms and not ruled by quantum mechanics" (Bell 1990, p. 16) is justified. In view of what we illustrated above, however, Bohr did believe that the classical measuring instrument is also made of atoms, for the simple reason that there could be no amplification process in experimental apparatuses without a physical interaction with quantum systems composed of atoms. In fact, given his entity realism (Folse 1986, Faye 1991), Bohr cannot be saddled with a wholesale simple-minded instrumentalism about unobservable entities (a charge that was explicitly leveled against Bohr by Popper (1963, pp. 130–160)).[11] The claim, to be illustrated later, that classical objects can, in appropriate measurement situations, be regarded as quantum objects renders Bell's charge wholly unjustified.

4 However, Bohr once explicitly denied that measurements generate physical outcomes: "phrases often found in the physical literature, as 'disturbance of phenomena by observation' or 'creation of physical attributes of objects by measurements' represent a use of words like 'phenomena' and 'observation' as well as 'attribute' and 'measurement' which is hardly compatible with common usage and practical definition and, therefore, apt to cause confusion" (Bohr 1998, p. 146). In other words, not only does the definite value of a certain kinematical or dynamical variable not exist prior to the measurement (which follows, explicitly after 1935, from his negation of Heisenberg's disturbance view) but, he seems to add, it is also not an effect caused in some sense by the apparatus and its interaction with the system. This quotation may seem in conflict with the physical view of measurement defended in the previous paragraph and, we argue, by Bohr himself.

In view of this, it is difficult to make sense of the last part of the above quotation until one realizes that to our knowledge this is the only place where Bohr defends theses like this. With these remarks he was probably objecting to claims made by some physicists that human consciousness plays an active role in determining the outcome of a measurement[12]: for Bohr, once we measure two particles in a singlet state in a Bell-type experiment, it is legitimate to claim that it is the interaction between one particle in one of the two wings of the experiment and the apparatus *A* that causes or "generates" the definite spin value by an amplification process involving the particle. The "creation" in the quotation must be interpreted, charitably, as referring to the necessary existence of a physical process of amplification that Bohr admitted and that we illustrated in the previous section. The result does not come "out of nowhere" or thanks to a subjective intervention of an observer: it is the process of amplification made possible by the apparatus that "creates" or generates the value being ascribed to the system, given the physical nature of the measurement argued for above.

---

[11] For a careful evaluation of Bohr's position toward instrumentalism, see among others Folse 1985; Howard 1994; Camilleri and Schlosshauer 2015; Zinkernagel 2015, and Dieks 2017.

[12] We want to thank Henry Folse for providing us with this reading.

In a word, we regard the first four points as sufficient to show that Bohr distinguished the avowedly *physical* nature of the process occurring in a measurement interaction – whose detailed description he abstained from giving for reasons that will become clear below—from the *product* of such a process, namely, the definite outcome, which must unavoidably be expressed in classical, communicable language.

This distinction, however, is not exempt from difficulties, which Bell raised with his characteristic acumen and which will help us to introduce the next section. Bohr's refusal to produce a constructive description of this process provoked Bell's charge as expressed in the famous paper "Against Measurement" (Bell 1990), where he claims that the word 'measurement' (together with others like 'microscopic', 'macroscopic', 'information', etc.) should not appear in a fundamental physical theory like quantum mechanics. "It would seem that the theory [quantum mechanics] is exclusively concerned about 'results of measurement' and has nothing to say about anything else. What exactly qualifies some physical systems to play the role of "measurer"? …If the theory is to apply to anything but highly idealized laboratory operations, are we not obliged to admit that more or less 'measurement-like' processes are going on more or less all the time, more or less everywhere? … The first charge against 'measurement', in the fundamental axioms of quantum mechanics, is that it anchors the shifty split of the world into 'system' and 'apparatus'" (Bell 1990, ibid.). [13]

As is well known, the 'shifty split' has been correctly regarded as the major difficulty raised by Bohr's approach to quantum mechanics. This difficulty can be reformulated by noting that, on the one hand, as we have shown, Bohr believed that measurement processes are real physical processes and, on the other, he defended in many passages of his work a skeptical attitude toward the claim that the wave function refers to a mind-independent reality, except as an encapsulation of probabilistic claims based on observed frequencies. However, the controversial question of whether Bohr's stress on 'measurement' can be defended by claiming that Bell's charge of the shifty split only depends on his *misrepresenting* Bohr as identifying 'macroscopic' with 'classical' and 'microscopic' with 'quantum', cannot be pursued further here. What we will now show is that such an identification ought to be reformulated in terms of the fact that Bohr consistently endorsed the distinction principle in an *epistemic* sense and the non-separability principle in an *ontic* sense at the same time.

[13] To this Bohr might have replied that none of the terms mentioned by Bell enter quantum theory, but they certainly enter our interpretation of that theory.

## 2. A remarkable tension between the two main tenets of Bohr's philosophy of quantum theory

Before beginning this new section, a caveat that partially applies also to the previous one seems appropriate. Any paper on Bohr's interpretation of quantum mechanics runs the risk of attributing him a single, unified view, not subject to change during his intellectual career.[14] Beller and Fine (1994), for instance, have argued that Bohr radically transformed his interpretation in order to adjust it to the EPR argument. We disagree, but a criticism of this view here would lead us astray and is fortunately unnecessary. First, if one investigates the scholarly literature on Bohr, scholars who have recognized some modifications in his view, such as Faye (1991), point out that the changes in Bohr's *arguments* before and after EPR were in any case motivated by an attempt to defend the *same* general interpretation of QM. A fuller justification of the substantial unity and continuity in Bohr's thought will become more evident from the tension in the two principles that we are going to discuss in what follows.

The *distinction principle* is perhaps the most distinctive feature of Bohr's philosophy of quantum theory, one that certainly remained a constant trait of his overall view: "…The essentially new feature in the analysis of quantum phenomena is, however, the introduction of a *fundamental distinction* between the measuring apparatus and the objects under investigation" (Bohr 1963, pp. 3–4, our emphasis). This distinction appears all the more controversial given that *also* Bohr believed that both consist of atoms, a fact that his critics, as we have seen, have not always appreciated. The distinction thesis can be interpreted in at least *three* different but related senses.

(i) In a first, *epistemic* sense, all empirical evidence must be reported by using readings of macroscopic apparatuses describable in the classical language, an *extension of the ordinary language* shared by the community of speakers. The intersubjectivity in question is necessary to avoid any reference to the observers' subjective mental states: "…by the word experiment we refer to a situation where we can tell others what we have done and what we have learned and that, therefore, the account of the experimental arrangements and the results of the observations must be expressed in unambiguous language with suitable application of the terminology of classical physics" (Bohr 1958, p. 39). Against erroneous interpretations given not only by Popper but also by Bunge (Faye 2017, p. 125), Bohr is not a subjectivist in the above sense, and conscious observers could in principle be replaced by inanimate physical devices. Ultimately, however, every observation must "be reduced to our sense perceptions" (Bohr 1934, p. 54), a statement that does not imply that the physical outcomes can be influenced by the intervention of the latter.

(ii) In a *semantic* sense, Bohr is pointing out that classical concepts like waves, particles, or in general, kinematic and dynamic notions cannot be applied to quantum physics *without restrictions*: "The quantum theory is characterized by the acknowledgment of a fundamental limitation in the classical physical ideas when applied to atomic phenomena. The situation thus created is of a peculiar nature, since our interpretation of the experimental material rests essentially upon the classical concepts" (Bohr 1927, p. 53). Relatedly, Landau and Lifschitz expressed the same concept by arguing that "quantum mechanics contains classical mechanics as a limiting case, yet at the same time it requires this limiting case for its own formulation" (Landau and Lifshitz 1977, p. 3).

[14] We thank one of the referees for suggesting that we stress this point.

(iii) In an *ontological* sense, however, according to Bohr there is *no ontic difference* between a classical (macroscopic) and a quantum (microscopic) entity (for this claim see also Dieks 2017, p. 314), since, as anticipated above, Bohr argued that quantum mechanics is *universal* and, depending on the experimental context, it also applies to the macroscopic realm typically described by the language of classical physics (Howard 1994; Landsman 2006, 2007; Camilleri and Schlosshauer 2015; Laudisa 2024). *Any macroscopic object therefore consists of quantum objects*. This is not in opposition to the already presented thesis that, in order to report the measurement results, we need an epistemic distinction between the classical and the quantum realm (Bächtold 2017).

After these three comments, it should be obvious why the distinction principle is at least *prima facie* in conflict with Bohr's other principle of non-separability, entailing a holistic view of the quantum measurement: in the famous paper concerning his discussion with Einstein, Bohr refers to the "*impossibility of any sharp separation between the behavior of atomic objects and the interaction with the measuring instruments which serve to define the conditions under which the phenomena appear*" (Bohr 1958, pp. 39–40, Bohr's emphasis). The basic notion of non-separability was already clear to Bohr when he first presented his ideas about complementarity at the Como conference in 1927. There he said: "The quantum postulate implies that any observation of atomic phenomena will involve an interaction with the agency of observation not to be neglected. Accordingly, an *independent reality* in the ordinary physical sense can neither be ascribed to the phenomena nor to the agencies of observation" (Bohr 1934, p. 54, our emphasis). In addition, the new "individuality of the atomic processes" that he also refers to in this paper is a direct consequence of the universal quantum of action (Bohr 1958, pp. 42-43), which stands for the indivisibility of the whole comprising the quantum system and the classical apparatus (see also Dieks 2017, p. 317).

The resulting tension with the distinction principle can be precisely expressed by the following, crucial question: *how can a macroscopic, classically describable entity A (the measuring apparatus) be distinct from a quantum entity S (a claim that presupposes some sort of "cut" or "split") if S does not possess a reality that is independent of A?* Here Bohr seems to be telling us explicitly that the only reality is the whole, or the non-separable state *S+A*. However, if we are correct in arguing that any separation between the quantum and the classical descriptions is necessary only for *epistemic reasons*, the distinction thesis just depends, contextually and holistically, on the given experimental setup, and as such it is essentially dictated by the particular aim of the experimenter.

Another way of describing this conflict consists in stressing that, as we saw above in section 1, Bohr refers to a genuinely physical correlation between a quantum system S and a classical apparatus *A* in which an amplification process occurs. Note that, on the one hand, by virtue of the meaning of 'amplify', the corresponding physical process of amplification seems to require the existence of *two entities*, one (the apparatus *A*) that amplifies the properties of the system *S* (the second entity) whose properties must be amplified. The existence of two entities presupposes the distinction principle. On the other hand, however, the interaction of *A* and *S* brings forth their physical non-separability.

How legitimate is it to label such non-separability as a form of *entanglement* in Schrödinger's (1935) sense as Howard (2021) maintains?[15] Although Schrödinger's term appeared in print only in 1935, he was aware of it already in 1926 and told Bohr about it in a letter of November 1926.[16] At that time, Schrödinger thought that it could not be explained (according to his own understanding of the wave function), but Jaksland's (2026) historical work suggests that Bohr in 1927 had already discovered that a measurement creates an "entangled" state between the object and the instrument in the sense of *a new form of nonseparable unity between a measuring instrument and a quantum system*. In a word, this non-separability was a well-known "phenomenon" in quantum mechanics before 1935 among all the leading physicists within the field, and even though Bohr never used the word entanglement, his view of a "new kind of individuality" whose parts are non-separable pointed to a very similar concept. From now on, we will use the word entanglement in relation to Bohr only in this precise sense.

The need for an epistemic distinction generated by the potential conflict between the distinction principle and the non-separability principle is further confirmed by his response to Einstein's thought experiment during the 1927 Solvay conference, when Einstein imagined an experimental set-up consisting of two diaphragms in front of a photographic plate. To our knowledge, the consequences of this response for clarifying Bohr's attitude toward the universal character of quantum mechanics have not been adequately considered.

The first diaphragm, a one-slit screen $D_1$, is either bolted to the lab table or suspended with springs and interacts with the particle *P* *before* its passage through the second diaphragm, a two-slit screen $D_2$. According to Einstein's argument, thanks to the conservation of momentum, if we consider the case where the first diaphragm is suspended, a downward or upward recoil of $D_1$ would enable us to determine precisely the trajectory of *P* going through $D_2$ (the former recoil causing the particle to go through the upper slit and the latter through the lower slit of $D_2$). We could therefore find out which of the two slits will be crossed (a position measurement), without destroying the interference pattern.[17]

Bohr replied to the thought experiment by claiming that *Heisenberg's relations* between position and momentum, and therefore the complete quantum mechanical description, had to be used *also for the suspended,*

---

[15] One of the referees correctly urged us to clarify this point.

[16] The history of the term 'entanglement' between 1926 and 1935 is well-documented by Jaksland (2026) https://philsci-archive.pitt.edu/30350/.

[17] Howard has claimed that with this thought experiment Einstein did not want to criticize Heisenberg's indeterminacy relation but rather stress the existence of non-local correlations between the two screens. Be that as it may, this interpretive issue does not affect our argument.

*classical and macroscopic screen* $D_1$. If this is the case, Bohr tells us, the precise measurement of the momentum of the recoil of the first screen, needed to determine the magnitude of its upward or downward motion, would entail a spread of its position and therefore an indeterminacy of the trajectory of the particle, thereby destroying the interference pattern (see Bohr 1958, p. 44). In a word, $D_1$ plus the springs should be regarded as a unique quantum object (Dorato 2017) and as such, during the measurement interaction, it would become entangled with the system to be measured.

From this thought experiment it seems reasonable to conclude that the *same* diaphragm (that is, the first one-hole screen $D_1$) can be described as a *classical* object when it is bolted to the lab table in the first description and as a quantum object when, in the second description, it is suspended with springs. The two descriptions of $D_1$ taken by itself are co-referential. Bohr's basic point is that these two descriptions refer to the same object $D_1$ only apparently, a fact that can be explained only by the *holistic*, *contextual* nature of the measurement setup expressed by his non-separability thesis. The crucial remark to keep in mind is that the same diaphragm $D_1$ is *used* by the experimenter in *two* different ways and therefore has a different function. Put it differently, we can do only one experiment at a time – a thesis that Bohr has insisted upon many times by later using the term complementarity also to refer to this sort of experimental limitation by moving from complementary magnitudes (spatiotemporal and causal) to complementary experimental situations. Strictly speaking, since the joint system '$D_1$ + $P$ + springs' is *not* identical to the quantum, holistic system '$D_1$ + $P$ + bolts', there is a sense in which the necessary existence of *two* different experiments implies that not only $P$ but *also* $D_1$ has an intrinsic, non-relational identity – and therefore has a separate existence—only when the two systems are not entangled. During the measurement interactions, $D_1$ has a purely *relational nature*, suggesting a perspectival view of quantum mechanics close to other recently defended views (Rovelli 1996, 1998, 2018, Dieks 2019, 2022, 2025, Dorato 2020, among others).

Notice that non-separability conjoined with contextuality has important consequences for the identity of the involved systems: separate existence requires the availability of identity criteria. For the sake of clarity, when $P$ and $D_1$ are not entangled, we could say that they are *materially separable*: $D_1$ has an identity that is clearly different from a quantum system $P$ when the two are *not* entangled. However, depending on its role or function in the experiment, $D_1$ becomes *functionally* non-separable from the measured system (and therefore entangled with it) when it is used in experimental situations similar to those described above.

More in detail, we shall say that two alleged entities are *materially separable* if and only if they, as kinds, fulfil different identity criteria. And, indeed, electrons (in virtue of their rest mass, spin and charge) and macroscopic objects (in our case *$D_1$ or $D_2$*) have different identity criteria. However, we shall also say that two alleged entities are *functionally non-separable* if and only if (1) they are materially separable, but (2) they do not have separate states satisfying different individuation criteria when they are entangled. *The functional non-separability depends on the particular role that the apparatus must play in the measurement context as determined by the kind of experiment that must be performed.*

In classical mechanics, material separability and functional separability go together in measurement, but this seems not to be the case when we are dealing with quantum measurements. Dealing with quantum systems, *two entities might be functionally non-separable but materially separable*. In particular, this would be the case whenever the atomic system and the apparatus interact during measurement and become a non-separable whole. By reporting what we already quoted above for the convenience of the reader, this fact explains why, during this process, and because of the quantum of action, according to Bohr "an independent reality in the ordinary physical sense can neither be ascribed to the phenomena nor to the agencies of observation" (Bohr 1934, p. 54). In a word, our suggestion is that by making a reasonable distinction between material separability and functional non-separability we may explain why, *prima facie*, mutually inconsistent assumptions in terms of the distinction principle and the non-separability principle do not pose a problem for Bohr. This has important implications for our treatment of, and proposed solution to, the conflict between the two principles.

The holistic approach just illustrated, and the necessity to do only one experiment at a time—a remark also used by Bohr in his response to the EPR argument (Bohr 1935)—explain and justifies[18] Bohr's pragmatist approach to the philosophy of quantum theory, where it is the changing aim of the experimenter that determines how to apply the distinction principle. We cannot say in advance, once and for all and for all possible experimental contexts, how to separate the classical from the quantum aspects of the phenomena, for the simple reason that in measurement interactions they are in principle always non-separable. Therefore, the distinction principle expresses the fact that, for any given circumstance, unless we select a classically describable apparatus out of the holistic, quantum character of the complete experimental situation, we cannot come to know the measurement outcome. Curiously enough, the main criticism levelled against Bohr has been motivated by his failing to provide a universally applicable, sharp distinction between the classical and quantum regime, one that Bohr thought could not be provided given the ontological holism that he defended and the consequent, irremediable contextuality: we submit that *it was because of the consequent impossibility of the objective existence of such a distinction when system and apparatus are entangled that there could not be a generally applicable, constructive theory of measurement.* As we will see in the next section, a principle-theory of measurement is a direct and most consistent consequence of the failure of functional separability.

## 3. Bohr's principle-theory view of the quantum measurement

We will now show that Bohr defended a *principle-theory* approach (in Einstein's sense) to the measurement problem and that as a consequence he could afford to reject any attempt to explain in physically detailed ways the process of amplification that, after the entanglement phase, leads to a definite measurement value. For this purpose, we need to recall briefly the principle/constructive theory distinction, which, unbeknownst to many, was originally proposed by Lorentz in terms of different attitudes toward scientific theories, and later adopted by Einstein in a famous 1919 article (Einstein 1954). Lorentz wrote: "General laws and principles" (thermodynamics, conservation laws) are 'prudent': by neglecting 'the inner constitution *of bodies,'* they apply to 'a wide variety of phenomena' but are 'only partly satisfactory'. Theories looking for 'mechanisms of the appearances' (statistical mechanics, Kelvin's vortex theory) broaden our understanding of particular physical systems but are 'more daring'" (Lorentz 1900c, p. 333, our emphasis, quoted in Frisch 2005, pp. 666–667). As is much better known, nineteen years later, Einstein used the same examples by referring to the special theory of relativity: "…the kinetic theory of gases seeks to reduce mechanical, thermal, and diffusional processes to movements of molecules – i.e., to build them up out of the hypothesis of molecular motion… The elements which form their basis and starting point [of principle theories] are not hypothetically constructed but empirically discovered ones, general characteristics of natural processes"… "the science of thermodynamics seeks by analytical means to deduce necessary conditions, [….] from the universally experienced fact that perpetual motion is impossible …." (Einstein 1954, p. 228).

[18] In the EPR case, the non-separability is between *two* quantum systems instead of one system and the apparatus. See Bacciagaluppi and Crull (2024).

The decisive passage in which Bohr refers to the non-indispensable feature of a detailed constructive account of the measurement process is in his 1949 piece written in honor of Einstein: "Notwithstanding all differences between the physical problems which have given rise to the development of relativity theory and quantum theory, respectively, a comparison of purely logical aspects of relativistic and complementary argumentation reveals striking similarities as regards the renunciation of the absolute significance of conventional physical attributes of objects. *Also, the neglect of the atomic constitution of the measuring instruments themselves, in the account of actual experience, is equally characteristic of the applications of relativity and quantum theory*" (Bohr 1949, p. 236; our emphasis).

The analogy with the special theory of relativity referred to by Bohr is two-fold. The first refers to the relativization of magnitudes achieved by the two theories: while in relativity, spatial and temporal intervals "taken by themselves", as Minkowski put it, become relative to an inertial frame and lose their Newtonian absoluteness, in quantum mechanics the existence of definite physical values of states that are not in eigenstates of observables becomes relative to the given experimental setup.

The second analogy, expressed by the emphasized sentence, is more important for our purpose. In Lorentz's dynamical explanation of the relativistic effects, contractions of length and temporal dilations are to be explained constructively and 'bottom up' by the inner constitution of the rulers and clocks. In relativity, they can be explained or at least understood geometrically in terms of projections from the structural constraints of a four-dimensional space-time that codifies the two principles of the theory, the relativity principle and the constancy of the velocity of light. Independently of whether a geometrical explanation of the relativistic effects is plausible, the second, stronger analogy is based on the fact that, in 1905, Einstein's choice to formulate the special theory of relativity as a principle theory was most probably suggested by his awareness of the *conflict* between the *discreteness* paradigm that he himself brought about by postulating the *discrete* nature of the electromagnetic radiation (light quanta) and the *continuity* implied by classical electromagnetism. In Bohr's case, an analogous tension between the distinction and the non-separability principle was resolved by adopting a principle-theory approach to the interaction between quantum systems *S* and apparatuses *A*.

In addition, note that there is a third analogy between the special theory of relativity and quantum mechanics as Bohr understood it: the quantum postulate and the light postulate contain different but universal constants of nature entering the most fundamental physical laws. By *constraining* what can be known of the physical world – *h* because of its discreteness, and *c* because of its finiteness – they are universal empirical generalizations or meta-laws that all phenomena must satisfy, as is the case in theories of principles.

According to Bohr, if quantum mechanics is regarded as a theory of principle – a thesis that he admittedly never defended explicitly by labeling in this way – we can and perhaps ought to neglect the details of the atomic constitution of the measuring instruments – which he considers to be real by virtue of his entity realism – *because* of the already introduced functional non-separability between the quantum system *S* and the apparatus *A*. As specified above, there is no non-arbitrary way to distinguish between *S* and *A* when they are functionally non-separable. The role of *h* here is key: given its discreteness, there is no way of saying whether it belongs to the system or the apparatus and therefore there is a sense in which it belongs to both. It is for this reason that the occurrence of a definite measurement outcome – say the position of a particle *S* – after an interaction between *S* and *A* depends *contextually* and *relationally* on the presence of a fluorescent screen entangled with *S*. It is the free choice of the experimenter to use a fluorescent screen *A* – whose function it is to reveal the location of the particle – that justifies the intervention of the distinction principle. The distinction between *S* and *A* is regained only *contextually* and *relationally* and *S* has a definite position on the screen only relative to the choice of that particular position-measurement apparatus.
In a word, in our view "neglecting the atomic constitution of *A"* and therefore avoiding a constructive account of the measurement interaction is a consequence of the ontic non-separability between *S* and *A*: the distinction (or split) between them is necessarily "shifty" and cannot be described in terms of their atomic constitution in the same way in all experimental cases. To summarize with one quotation what we have been writing above, according to Bohr "the concept of observation is in so far arbitrary as it depends upon which objects are included in the system to be observed" (Bohr 1958, p. 54) which is an obvious expression of the contextuality and relationality of the result of any quantum experiment.

## 4. Comparing three attempts to solve the conflict between the two principles

On the premises that Bohr was (i) in favor of a single-world approach to quantum mechanics, (ii) regarded any classical object as treatable as a quantum object, (iii) defended a physical theory of the quantum measurement but (iv) refused to attribute to the word 'collapse' any physical status, we will now discuss three possible ways of understanding his position about the disentanglement between the quantum and the classical and thereby shed light on the conflict between the two principles illustrated above. We will examine in turn Howard's (2021), Zinkernagel's (2016), and Dieks's (2017) proposal in order to argue that the view defended here is more convincing than these three.

### 4.1. Howard's solution of the distinction problem in terms of decoherence

Howard has also correctly given due emphasis to the tension between the distinction and the non-separability principles: "On the one hand, entanglement is a universal fact about interacting systems as described by quantum mechanics, meaning that the instrument and the object form an indissoluble whole … On the other hand, objectivity requires the use of classical modes of description in accounting for measurements and their outcomes. How is this dialectical tension to be resolved?" (Howard 2021, p. 163). In this sub-section we will raise three criticisms against Howard's proposal to solve the conflict. The first criticism C1 involves Howard's claim that, according to Bohr, *the complementarity principle follows from entanglement.* The second criticism C2 stresses the fact that Bohr, as we argued in the section above, does not invoke, and need not invoke, a dynamical explanation of the emergence of the experimental outcomes. The third, more serious criticism C3 involves his claim that decoherence entails that the classical world is to be regarded as an "emergent simulacrum" (in his words) of the quantum world.

C1. In order to support his decoherence-based account of Bohr's approach to the relation between classicality and the quantum, Howard stresses (as we did) Bohr's early recognition of the centrality of the entanglement of systems and apparatuses. In his opinion, the tension between the two principles derives from the fact that complementarity is a consequence of entanglement: "There is widespread misunderstanding of Bohr's doctrine of complementarity, much of it owing not to Bohr's alleged obscurity, but to a failure to understand the extent to which and manner in which complementarity was seen by Bohr to be a direct logical consequence of the physical fact of entanglement" (Howard 2021, p. 154). According to Howard, to determine which slit of D2 the particle passed through (recall the discussion above about the position measurement), we must treat the first screen as a quantum object subject to the indeterminacy relation. While we agree with this point, it seems that one can have an entangled system independently of a measurement apparatus. Take Schrödinger's cat. If quantum mechanics is complete and the evolution is always linear, the cat is entangled with the radioactive particle, the flask, the venom, etc., but there seems *no* complementarity between conjugate variables: the latter may follow from entanglement only in particular experimental situations. In the Como paper, Bohr's concept of complementarity follows from Heisenberg's uncertainty relations together with the indispensable use of classical concepts for the description of experimental outcomes, not just from the physical entanglement between the object under investigation and the apparatus.

C2. Bohr does not invoke and need not invoke a dynamical account of the definiteness of the experimental outcome. Although decoherence was first introduced into quantum mechanics after Bohr's death and in the context of the many-worlds interpretation, Camilleri and Schlosshauer (2015) and Schlosshauer and Camilleri (2017) have argued that this notion provides Bohr's epistemological approach with a *dynamical theory* of the uncontrollable interaction between the measurement apparatus and the system under investigation. Thus, they conclude that the two notions, complementarity and decoherence, are not in conflict. In addition, Howard believes that not only can these two concepts coexist consistently but also that decoherence can explain the conflict between our two principles by invoking entanglement. Howard claims that the decisive passage favoring his reading is the following remark by Bohr: "In the system to which the quantum mechanical formalism is applied, it is of course possible to include any intermediate auxiliary agency employed in the measuring process. Since, however, all those properties of such agencies which, according to the aim of the measurements, must be compared with the corresponding properties of the object, must be described on classical lines, their quantum mechanical treatment will for this purpose be essentially equivalent with a classical description" (Bohr 1998, p. 104). On the basis of this passage, Howard interprets Bohr as holding that even if we had a more detailed account of the measurement process in quantum mechanical terms, for all practical purposes – that is, in Bell's expression, FAPP – such an account would be redundant.

However, we have shown before that Bohr can solve the distinction problem by stressing that what is to be considered as classical depends on the variable and contextual aim of the experimenter, so that the FAPP but *constructive* "solution" of the measurement problem offered by decoherence is not needed. Decoherence requires one to consider the wave function as representing a real physical evolution of the system under investigation, whereas Bohr considered it to be symbolic and therefore as a means for predicting the statistical outcome of measurements. Hence, we are skeptical of Howard's claim that "environment-induced decoherence (which is not really collapse, but only a semblance thereof) was, all along, the real point toward which Bohr was gesturing with the doctrines of complementarity and classical concepts" (Howard 2021, pp. 170-171).

C3. Our third, stronger criticism, shows that Bohr never thought of the classical realm as "a simulacrum" emerging from the quantum world since if the former is somewhat grounded in the latter, the former has an ontological autonomy justified by our use of the ordinary language and the categories that we employ to structure the world of our experience. To the extent that the separation between the quantum and the classical is somewhat arbitrary and context-dependent because it cannot be established once and for all, a constructive account of the measurement process is not needed. Here Bohr's view could plausibly be read in terms of Sellars' (1963) well-known distinction between the manifest and the scientific image of the world: "to complete the scientific image we need to enrich it not with more ways of saying what is the case, but with the language of community and individual intentions, so that by construing the actions we intend to do and the circumstances in which we intend to do them in scientific terms, we directly relate the world as conceived by scientific theory to our purposes, and make it our world and no longer an alien appendage to the world in which we do our living" (Sellars 1963, p. 40). That the classical world is a simulacrum would require the classical world to consist of mere appearances, but this view fits badly with Bohr's insistence on the indispensability of the use of classical concepts to understand not only the experimental results but the entire experimental practice.

Bohr's appeal to common language is needed to understand the physical practice and how *human intentions* play a role in determining when the different kinds of experiments should be described classically or quantum mechanically. This is the main sense in which we should understand Bohr's pragmatism. Equally important is Bohr's understanding of objectivity as what can be communicated unambiguously and therefore intersubjectively. These two claims together show that Bohr thought that the preconditions for a rational and objective description of the world are exactly those features that Sellars regarded as essential for the characterization of the manifest image. In addition, Howard points out that in order to argue for the absence of collapse in Bohr's treatment of quantum mechanics, the density matrix can be regarded as a classical mixture whose components become separated after a measurement. However, there is no need to *explain* why Bohr considered collapse as non-existent since, according to him, the wave function must be regarded as symbolic in all circumstances for the reasons explained above. Moreover, according to Bohr, "derived real functions like densities and currents are only to be regarded as expressing the probabilities for the occurrence of individual events observable under well-defined experimental conditions"[19] (Bohr 1998, p. 144).

### 4.2. Zinkernagel's account of the tension between the distinction principle and the non-separability principle

The importance that Bohr assigned to the holistic character of quantum mechanics can be gathered from the fact that it enters into the definition given to the main physical principle that he defended, namely the *quantum postulate*, which he regarded as the essence of the quantum theory, and which "attributes to any atomic process an essential discontinuity, or rather individuality, completely foreign to the classical theories and *symbolized* by Planck's quantum of action" (Bohr 1934, p. 53, our emphasis).[20] The use of the word 'individuality', pointing to the fact that parts of the entangled systems have no separate existence, must be further clarified. Behind Bohr's reference to this word there is in fact an additional motivation that he made explicit in the last part of the quotation above. With its discrete, indivisible nature, Planck's constant *h symbolizes* a new form of 'existence' (the entanglement between the agents of observations *A* and the systems *S*) because it represents a new form of atomism, namely *the atomization of action*.[21]

---

[19] Here Bohr is presumably referring to density matrices and probability currents.
[20] Dieks agrees with our remark here (2017, p. 314).
[21] We owe this expression to Massimo Pauri.

Given the three controversial arguments that Bohr lists to deny any reality status to the wave function – its complex nature, the fact that it evolves in a high-dimensional configuration space, and its non-relativistic character— it is important to discuss a viewpoint expressed by Henrik Zinkernagel that is close to our own but diverges from it in an important respect[22]: "We can . . . say that, for Bohr, the collapse is not physical in the sense of a physical wave (or something else) collapsing at a point. But it is a *description*— in fact the best or most complete description of something happening, namely the formation of a measurement record" (Zinkernagel 2016, p. 14, our emphasis). While we agree with Zinkernagel that the measurement for Bohr *is* a physical process (see section 1 above), given Bohr's conviction that the wave function cannot be given an ontological interpretation, it seems to us that using the word 'collapse' in relation to Bohr's view might be misleading. In fact, if Bohr had thought that some type of collapse had a role to play in understanding quantum mechanics, in some of his published papers he would have mentioned this notion.

This might be only a semantic slip. In the same paper, however, and more seriously, following Landau and Lifshitz (1977), Zinkernagel has furthermore tried to offer a universally applicable account of the transition from the superposed 'and' to the exclusive 'or' in measurement interactions (Zinkernagel 2016). His key idea lies in the claim that in any such interaction, the quantum system *S* is entangled *only with a part* of the macroscopic apparatus *A*. But this solution is unsatisfactory, since the distinction between a part *P* of *A* and *A* itself is unavoidably vague and indefinite ("How large must the part be?") and generates the same problem that Bohr tried to avoid. Introducing parts *P* of a macroscopic apparatus *A* as entities of intermediate size between quantum systems and classical apparatuses could only be justified by abandoning the non-separability principle, which for Bohr, as argued before, is *essential*. If *A* is entangled with *S*, both systems have no independent existence, and have a non-intrinsic, relational identity because they are functionally non-separable. A fortiori, parts of A have the same characteristics. In other words, A cannot be described as the sum of its parts *P*, because *S* + *A* form an indivisible, functionally and ontologically non-separable whole.

Another problem raised by partial entanglement is that the observer selects which kind of property she wants to measure by choosing a specific kind of measuring apparatus (a fixed or a movable diaphragm as in the example above). As we have already seen, in the Nature-version of the Como-paper, Bohr (1928), he noted that in quantum mechanics it is the observer that determines which parts of the arrangement "are included in the system to be observed". This implies that by making her choice among the possible experiments, the observer determines which parts are to be regarded as classical and which not as a function of what she wants to find out, with the proviso that, as Bohr has emphasized: "it is certainly not possible for the observer to influence the events which may appear under the conditions [s]he has arranged" (Bohr 1958, p. 51)

---

[22] As many interpreters have noted, of these three arguments, only the second seems solid.

### 4.3. Dieks' proposal[23]

In his discussion of measurement and complementarity, Dieks (2017) points to Bohr's idiosyncratic use of the term 'individuality'. Bohr used this term to refer to the situation where the system and the measurement apparatus during the interaction form a united whole due to the quantum of action. Even though Bohr talked about *two* entities interacting in an unanalyzable way, he also drew the conclusion that they were forming a united whole. However, Dieks argues that since classical concepts for Bohr are indispensable for describing macroscopic objects, it immediately follows that measurements have only one single outcome: "Bohr's interpretation does not face the 'measurement problem' in the form in which it is often posed in the foundational literature; for Bohr *this* [the uniqueness of the outcome] *is not something to be explained, but rather something that is given and has to be assumed to start with. It is a primitive datum … So, the measuring problem in its usual form does not exist; it is dissolved" (Dieks 2017*, p. 319, our emphasis). This line of argument is close to the one we defended in the third section and is therefore important to illustrate it in more detail.

First of all, we should point out that Dieks is more sceptical than we and Howard are about Bohr's awareness of the significance of entanglement, although he recognizes that Bohr's term "individuality" "actually corresponds to entanglement in the formalism" (ibid., p. 324). Secondly, he agrees with the claim illustrated above that Bohr's view of the quantum formalism does not count as instrumentalism, by reporting correctly that Bohr's stressing of the 'symbolic nature of the wave function' just connects to the fact that it does not provide us with a *visual representation*. He adds that Bohr took the formalism seriously in the sense that he held that the uncertainty relations and the canonical commutation relations were the formal preconditions for complementarity. Hence, since the principle of decoherence and complementarity are not mutually inconsistent, one may use the formalization of decoherence as a representation of the interaction in which we can describe one component of the system classically because the relevant "latitude" in Bohr's sense becomes negligible[24]. But, as Dieks notices, with no collapse (which was Bohr's view), after the correlation has taken place, the unitary description of the interaction of the object and the measuring device always produces an entangled state. Consequently, as argued before, decoherence cannot solve the measurement problem except as a FAPP move.

Thirdly, however, Dieks does not think that the preservation of entangled states is a challenge for Bohr's view because such a view does not begin with the quantum formalism but rather with the supposition that the mathematical symbols get their meaning from their identification with physical quantities on the macroscopic level and therefore from our use of ordinary language and classical terms. Hence, Dieks argues, the uniqueness of experimental results is not something that Bohr believed had to be explained, but, as affirmed in the above quotation, something which we must accept as *a brute fact*, a view that is close to the one that we defended in the third section, but that we justify in a different and, we believe, more convincing way, given that our appeal to a methodological commitment to a principle theory of measurement can be used as a *justification* or explanation of Dieks' "brute-fact" view. It is the very non-separability between classical and quantum that suggests a contextual solution of the measurement problem in the sense explained above.

---

[23] Here it is essential to note that, given the context, we limit our discussion to this paper, without mentioning the later 'relationist' development of Dieks' view noted above.

[24] Bohr talks about classical descriptions applying only "within definite latitudes" — meaning that when we use classical concepts (like position or momentum) to describe a quantum object or a measuring instrument, the classical descriptions are not exact but valid only within a certain approximation (latitude) or a permitted range of indefiniteness.

According to Dieks, Howard assumes that Bohr treated the unitary interaction between the object, the measuring device, and the environment *as if* after the interaction they are not entangled. As he argues, the fact that the superposition between the object and the instrument continues after measurement "... suggests that the full superposition by itself cannot have the desired meaning, and that Bohr therefore had to have recourse to a sleight of hand." (Dieks 2017, p. 327). Dieks himself maintains that for Bohr "superposition does not have an empirical meaning independently of its interpretation via classically described experiments, so no replacement by another state is needed" (ibid.). His conclusion therefore is that Bohr holds a non-collapse interpretation, a view on which, however, both we and Howard agree.

Following up on the relational character of Bohr's interpretation, Dieks argues further to the effect that the distinction thesis and the non-separability thesis are context-dependent and perspectival (Dieks 2018, 2022, 2025) because an instrument *A* in a hermetically closed chamber that is interacting with a superposed quantum system *S* will register a unique result even though, with respect to an outside observer, the joint systems *A+S*, according to the unitary evolution, are entangled.[25] However, Bohr would probably have added to this well-known formulation of Wigner's friend argument the remark that the wave function "is to be considered as a *tool* for deriving predictions of definite or *statistical character*" (see above) and even though he acknowledges that, by measuring one entangled particle in one wing of the experiment, we can predict the corresponding value of the other. If we measure entangled states, there is something empirically real that we measure that, however, is not represented by the wave function.

In the end, according to all three interpretations discussed here, the wave function is, for Bohr, nothing more than a measure of the statistical outcomes of measurements.

---

[25] Recall that we stressed the relational nature of the quantum system when we described the thought experiment with the two diaphragms.

## 5. Wrapping up

It is appropriate to conclude by comparing Howard's, Zinkernagel's, and Dieks's views with the view offered here as a solution to the distinction problem. By criticizing Howard's constructive approach in its ramifications, we have argued that his proposal is difficult to sustain because of his dynamical reading of the role of wave function and his understanding of the classical world as a simulacrum. Zinkernagel's claim that Bohr assigns the word 'collapse' a denoting role runs against textual evidence, and his version of the measurement problem as involving entanglement only with a part of the classical apparatus neglects the holistic, quantum nature of the apparatus. The difference between our view and Dieks' explanation of the conflict is more nuanced. What the latter calls a 'brute fact' for us can in *some sense* be justified or even explained by a methodological appeal to regard quantum mechanics as "a theory of principle", a claim that enables us to refrain – at least temporarily – from describing the emergence of a definite outcome in a constructive way.

Thus, in contrast to the three mentioned authors, we think that Bohr's view of measurement in quantum mechanics can best and most fruitfully be understood by Einstein's distinction between theories based on empirically discovered principles and theories based on hypothetical constructions. On the one hand, Bohr assumed that a quantum system and an instrument are ontically non-separable during a particular measurement process, in such a way that they do not have independent existence and that the measured value therefore does not exist prior to the measurement. In this sense, he can be portrayed as a quantum universalist. On the other hand, he argued that one must distinguish between system and apparatus if one wants to speak meaningfully of a measurement: the distinction principle holds epistemically. The problem for Bohr was that these two assumptions seemingly are in conflict since the epistemology is not aligned with his ontology. The appeal to the contextuality of the split, and therefore to a principle theory of measurement, was his way of resolving the conflict. We argued that Bohr viewed measurement as a physical process, but we also introduced a distinction between what we called material separation and functional separation. This distinction enables us to say that although system and apparatus are functionally inseparable during the measurement itself and therefore cannot be described constructively, we can still speak of a material separation between them, which permits us to refer to an objective measurement of one of the system's observables.